\documentclass[%
prd,
nofootinbib,
superscriptaddress,
dvipdfmx
]{revtex4}
\usepackage{amsmath,amssymb,bm,float,mathrsfs}
\usepackage{graphicx}
\usepackage{dcolumn}
\usepackage{color}
\usepackage{hyperref}
\usepackage{comment}

\usepackage{ulem}

  \newcommand{\beq}{\begin{equation}}
  \newcommand{\eeq}{\end{equation}}
  \newcommand{\al}[1]{\begin{align} #1 \end{align}}
  \newcommand{\bi}{\begin{itemize}}
  \newcommand{\ei}{\end{itemize}}
  \newcommand{\bc}{\begin{center}}
  \newcommand{\ec}{\end{center}}
  \newcommand{\GG}{\xi}
  
    \newcommand{\SY}{\textcolor{red}}
  
  \def\dd{\mathrm{d}}
  
  \def\pd{\partial}
  
  \newcommand{\ave}[1]{\left\langle #1 \right\rangle}

\begin{document}

\begin{flushright}
RUP-17-17
\end{flushright}

%****************************** TITLE **********************************%
\title{Constraining modified theory of gravity with galaxy bispectrum}

%***************************** AUTHOURS ********************************%

\author{Daisuke Yamauchi}
\email[Email: ]{yamauchi"at"jindai.jp}
\affiliation{
Faculty of Engineering, Kanagawa University, Kanagawa, 221-8686, Japan
}
\author{Shuichiro Yokoyama}
\affiliation{
Department of Physics, Rikkyo University, Tokyo 171-8501, Japan
}
\affiliation{
Kavli IPMU (WPI), UTIAS, The University of Tokyo,
Kashiwa, Chiba 277-8583, Japan
}
\author{Hiroyuki Tashiro}
\affiliation{
Department of Physics, Graduate School of Science, Nagoya University, Aichi 464-8602, Japan
}

%****************************** ABSTRACT *******************************%
\begin{abstract}
We explore the use of galaxy bispectrum induced by the nonlinear gravitational evolution 
as a possible probe to test general scalar-tensor theories with second-order equations of motion.
We find that time dependence of the leading second-order kernel is approximately
characterized by one parameter, {\it the second-order index,} which 
is expected to trace the higher-order growth history of the Universe.
We show that our new parameter can significantly carry new information about the non-linear growth of structure. 
We forecast future constraints on the second-order index as well as the equation-of-state parameter
and the growth index.
\end{abstract}

\pacs{}
%***************************** PREPRINT ********************************%
\preprint{}

\maketitle
%%%%%%%%%%%%%%%%%%%%%%%%%%%%%%%%%%%%%%%%%%%%%%%%%%%%%%%%%%%%%%%%%%%%%%

%%%%%%%%%%%%%%%%%%%%%%%%%%%%%%%%%%%%%%%%%%%%%%%%%%%%%%%%%%%%%%%%%%%%%%%%%
%=======================================================================%
\section{Introduction}
%=======================================================================%
%%%%%%%%%%%%%%%%%%%%%%%%%%%%%%%%%%%%%%%%%%%%%%%%%%%%%%%%%%%%%%%%%%%%%%%%%

It is one of the biggest challenges of modern cosmology to understand the physical origin of
the present cosmic acceleration of the Universe.
The origin of the cosmic acceleration is expected to be connected to fundamental theory 
beyond the current standard model of particle physics.
It might eventually require the presence of a new type of energy, usually called dark energy.
As another possibility, the accelerated expansion might arise due to a modification of general relativity~(GR)
on cosmological scales.
A variety of theoretical scenarios have been proposed in literature and carefully compared with observational data
(see Refs.~\cite{Tsujikawa:2010zza,Clifton:2011jh,Koyama:2015vza} for reviews).

Among many varieties of current cosmological observational data, 
measuring the growth rate of the density fluctuations, $f(a)$, is believed to be a powerful tool to test the nature of the dark energy
or the modification of the theory of gravity responsible for the present cosmic acceleration.
%In recent years, 
the growth rate of large-scale structure is mainly measured by observing galaxy peculiar velocities along the line of sight
through redshift-space distortion (RSD)
measurements 
\cite{Kaiser:1987qv,Hamilton:1992zz}.
To compare the observational data and theoretical predictions efficiently, 
it should be useful to introduce a phenomenological parameter. 
A minimal approach to test the theory of gravity from the measurement of the growth rate of large-scale structure 
is to introduce an additional
parameter called {\it gravitational growth index},
$\gamma$, defined through the growth rate~\cite{Linder:2005in,Linder:2007hg}:
\al{
	f(a)=\widetilde\Omega_{\rm m}^\gamma (a)
	\,,
}
where $\widetilde\Omega_{\rm m}(a)$ denotes the matter density fraction of
the total energy density at a cosmic scale factor $a$.
In the standard cosmological model responsible for the present cosmic acceleration, called $\Lambda$ cold dark matter ($\Lambda$CDM) model with GR,
we expect the growth index to be approximately constant with $\gamma\approx 0.545$.
Although the current constraints on the growth index have been reported~\cite{Mueller:2016kpu,Sanchez:2016sas,Grieb:2016uuo}, 
at the moment, there is no evidence for a departure from the standard $\Lambda$CDM model.
However, since there are numerous different ways of modeling the
landscape of cosmological models,
it is further required
to consider new possible parametrizations as the signature of the modified gravity theory.

In this paper, we focus on the {\it quasi-}nonlinearity of the growth of large-scale structure
as a way to provide new insight into the modified theory of gravity.
As an observable for such a nonlinearity, we investigate
the bispectrum of the biased object such as a
dark matter halo or galaxy 
which are frequently discussed as 
a useful tool to constrain the higher-order statistical
nature of cosmological perturbations~(see {\it e.g.}, Ref.~\cite{Yamauchi:2016wuc}
for constraining non-local types of primordial non-Gaussianity).
Even if the primordial perturbations are Gaussian,
the non-zero halo/galaxy bispectrum should be generated
from the late-time nonlinear gravitational evolution
of the density fluctuations and such a nonlinearity should have new information
about the modification of the gravity theory, which would not be
imprinted on the growth index in the linear perturbation theory.
As examples, Refs.~\cite{Takushima:2013foa,Takushima:2015iha}
have discussed the bispectrum of the matter density fluctuations
in Horndeski theory which has been known as a most general scalar-tensor theory with
the second-order equations of motion, and they have shown that the deviation from the standard $\Lambda$CDM model with GR
can be included in a time-evolving coefficient in the kernels of the second-order density perturbations denoted by $\lambda(a)$.

Here, we study the possibility of future planned galaxy surveys for providing
significant information on the modification of gravity theory responsible for the present accelerated expansion of the Universe, based on the
matter bispectrum formula derived in Refs.~\cite{Takushima:2013foa,Takushima:2015iha}.
For this purpose, we first
propose a new useful parametrization of the second-order perturbative kernel:
\al{
	\lambda (a)=\widetilde\Omega_{\rm m}^\GG (a)
	\,.
}
Here $\GG$ is our new parameter, {\it the second-order index}, to trace the nonlinear growth history, encompassing deviations in
the wide theoretical framework.
Actually, there are models in which the expansion history and the growth
rate in the linear perturbation theory are same in the fiducial model
but the different value of $\GG$ can be obtained.
We  show that precise measurement of RSDs by future galaxy surveys can distinguish and hopefully exclude 
these cosmological models.

The paper is organized as follows.
In Sec.~\ref{sec:galaxy bispectrum in redshift space}, we first give the basic equations for the galaxy bispectrum
in redshift space.
In Sec.~\ref{sec:Horndeski theory of gravity},
following Refs.~\cite{Takushima:2013foa,Takushima:2015iha},
we review the matter bispectrum in the Horndeski theory of gravity, the most general scalar-tensor theory
with second-order equations of motion.
In Sec. \ref{sec:analytic evaluation},
we estimate the asymptotic values
of $\gamma$ and $\GG$ by introducing the effective field theory parameters which can make the understanding of the signal
of the modification of gravity theory easier.
In Sec.~\ref{sec:Fisher analysis}, to see the impact of the existence of the new parameter on future galaxy surveys, 
based on the Fisher analysis, we numerically estimate the expected constraints on $\gamma$
and $\GG$ as well as the equation-of-state parameters.
Finally, Sec.~\ref{sec:conclusion} is devoted to summary and conclusion.

%%%%%%%%%%%%%%%%%%%%%%%%%%%%%%%%%%%%%%%%%%%%%%%%%%%%%%%%%%%%%%%%%%%%%%%%%
%=======================================================================%
\section{Galaxy bispectrum in redshift space}
\label{sec:galaxy bispectrum in redshift space}
%=======================================================================%
%%%%%%%%%%%%%%%%%%%%%%%%%%%%%%%%%%%%%%%%%%%%%%%%%%%%%%%%%%%%%%%%%%%%%%%%%

Galaxy redshift survey provides a map of galaxies in redshift space.
In this map, the radial position of a galaxy is given by the observed
radial component of its relative velocity to an observer, which is a combination of the Hubble
recession and the peculiar velocity.
Therefore, the mapping from a position~$\bm x$ of a galaxy in real space
to a position~$\bm s$
in redshift space is described as
\al{
{\bm s} = {\bm x} + {v_z(a, {\bm x}) \over aH} \hat {\bm z}
	\,,
}
where $v_z(a,{\bm x})$ represents a peculiar velocity of the underlying matter density field along the line-of-sight (we take the line-of-sight direction
to be $\hat{\bm z}$), $H$
is the Hubble parameter, and $a$ is the cosmic scale factor which 
we use as a time coordinate.
This mapping gives us the conversion from the density contrast in real
space, $\delta$, to that in redshift space, $\delta_s$.
Since it is useful to investigate
the spatial density distribution in Fourier space, up to the second order in the perturbation theory, we provide 
the density contrast in redshift space as 
the Fourier component~\cite{Scoccimarro:1999ed}:
\al{
	\delta_s(a,{\bm k})
		=&\int\dd^3{x}\, e^{-i{\bm k}\cdot{\bm x}}e^{-ik_zv_z/(aH)}\bigl[1 + \delta (a,{\bm x})\bigr] - \delta^{3}_{\rm D}({\bm k})
	\notag\\
		=&\bigl[\delta (a,{\bm k})+\mu^2 \theta (a,{\bm k})\bigr] \notag\\
		  &+\int {\dd^3{k}_1\dd^3{k}_2 \over (2 \pi)^3 }\delta_{\rm D}^{3}({\bm k}_1+{\bm k}_2-{\bm k})
			\left[\delta (a,{\bm k}_1)\,
			k\mu\frac{\mu_2}{k_2}\theta (a,{\bm k}_2) + {k^2 \mu^2 \over 2} {\mu_1 \over k_1}  {\mu_2 \over k_2} 
			 \theta(a,{\bm k}_1)\theta(a,{\bm k}_2) \right]
			+\cdots
	\,,
}
where $\delta^{3}_{\rm D}$ is a 3-dimensional Dirac's delta function,
$\mu$ is the cosine of the
angle between the wave-vector $\bm k$ and the line-of-sight
direction~$\hat{\bm z}$~(similarly, $\mu_i$ is the cosine for ${\bm k}_i$
with $i=1,2$).
Here 
$\delta (a,{\bm k})$ and $\theta(a,{\bm k})$ are Fourier components of
the density contrast and the scalar velocity divergence, $\theta (a,{\bm x}) := - \nabla \cdot {\bm v}(a, {\bm x})/(aH)$ in real space, respectively.
For the pressureless non-relativistic matter such as CDM and baryons, the evolution equation for the linear density fluctuations, $\delta$,
does not depend on the wavenumber and hence we can express the time dependence of the matter density fluctuations in the linear theory
independent with the wavenumber as
\begin{equation}
\delta (a, {\bm k}) = D_+(a) \delta_{\rm L} ({\bm k}),
\end{equation}
where $D_+ (a)$ is a growth factor of the growing mode in the linear theory and $\delta_{\rm L}({\bm k})$ is the random initial density
perturbations. Furthermore, by using the continuity equation, the scalar velocity divergence, $\theta$, is given by the logarithmic time derivative of
matter density fluctuations as
\begin{equation}
\theta (a, {\bm k}) =  f(a) \delta (a,{\bm k}),
\end{equation}
where $f(a) := \dd\ln D_+/\dd\ln a$ is called a linear growth rate.
By using above expressions, the expansion forms of $\delta(a, {\bm k})$ and $\theta (a, {\bm k})$ are respectively given as
\al{
	&\delta (a,{\bm k})
		=D_+(a)\delta_{\rm L}({\bm k})
			+D_+(a)^2\int\frac{\dd^3{k}_1\dd^3{k}_2}{(2\pi )^3}\delta_{\rm D}^3({\bm k}_1+{\bm k}_2-{\bm k})F_2({\bm k}_1,{\bm k}_2;a)\delta_{\rm L}({\bm k}_1)\delta_{\rm L}({\bm k}_2)
			+\cdots
	\, \label{eq:def_kernel_F} ,\\
	&\theta (a,{\bm k})
		=f(a)D_+(a)\delta_{\rm L}({\bm k})
			+f(a)D_+^2(a)\int\frac{\dd^3{k}_1\dd^3{k}_2}{(2\pi )^3}\delta_{\rm D}^3({\bm k}_1+{\bm k}_2-{\bm k})G_2({\bm k}_1,{\bm k}_2;a)\delta_{\rm L}({\bm k}_1)\delta_{\rm L}({\bm k}_2)
			+\cdots
	\,,
\label{eq:def_kernel_G}
	}
where 
the functions $F_n$ and $G_n$ are the so-called perturbative kernels.

Our interested observable is 
the galaxy distribution, not the matter density distribution.
Here, for simplicity, we introduce a local, nonlinear bias model for galaxies, 
in which the galaxy density contrast
can be expanded as a Taylor series of the underlying dark matter density contrast,
\al{
	\delta_{{\rm gal}}(a,{\bm x})=b_1\delta (a, {\bm x}) +\frac{1}{2}b_2\delta^2 (a, {\bm x})+\cdots
	\,,
}
with the bias parameters $b_i$.

Then, we can obtain the galaxy density contrast $\delta_{{\rm
gal},s}(a, {\bm k})$ in redshift
space as
\al{
	\delta_{{\rm gal},s}(a,{\bm k})
		=D_+(a)Z_1({\bm k}\,;a)\delta_{\rm L}({\bm k})
			+D_+^2(a)\int\frac{\dd^3{\bm k}_1\dd^3{\bm k}_2}{(2\pi )^3}\delta_{\rm D}^3({\bm k}_1+{\bm k}_2-{\bm k})Z_2({\bm k}_1,{\bm k}_2;a)\delta_{\rm L}({\bm k}_1)\delta_{\rm L}({\bm k}_2)
			+\cdots
	\,. \label{eq:delta_gal_s}
}
Here 
the linear- and second-order RSD kernels, $Z_i$, are defined as~\cite{Scoccimarro:1999ed}
\al{
	&Z_1({\bm k}\,;a)=b_1+f\mu^2
	\,,\\
	&Z_2({\bm k}_1,{\bm k}_2;a)
		=b_1F_2({\bm k}_1,{\bm k}_2;a)
			+f\mu_{12}^2 G_2({\bm k}_1,{\bm k}_2;a)
			+\frac{fk_{12}\mu_{12}}{2}\biggl[\frac{\mu_1}{k_1}\left( b_1+f\mu_2^2\right) +\frac{\mu_2}{k_2}\left( b_1+f\mu_1^2\right)\biggr]
			+\frac{1}{2}b_2 \label{eq:sokernel}
	\,,
}
where 
$\mu_{12}={\bm k}_{12}\cdot\hat{\bm z}/k_{12}$ with ${\bm k}_{12}={\bm k}_1+{\bm k}_2$\,.

Eq.~(\ref{eq:delta_gal_s}) allows us to calculate the power spectrum and
bispectrum of galaxies in redshift space.
For simplicity we assume that 
the initial density field $\delta_{\rm L}({\bm k})$ obeys the Gaussian random distribution with
\al{
	\ave{\delta_{\rm L}({\bm k}_1)\delta_{\rm L}({\bm k}_2)}=(2 \pi)^3P_{\rm L}(k_1)\delta^3_{\rm D}({\bm k}_1+{\bm k}_2)
	\,.
}
In the leading order of the perturbations, 
the power spectrum of the redshift-space galaxy
is simply described by 
\al{
\ave{\delta_{{\rm gal},s}(a,{\bm k}_1)\delta_{{\rm gal},s}(a,{\bm k}_2)} = (2 \pi)^3P_s ({\bm k}_1;a)\,\delta^3_{\rm D}({\bm k}_1+{\bm k}_2)
	\,,
}
with
\al{
	P_s({\bm k}\,;a)=D_+^2(a)Z_1^2({\bm k}\,;a)P_{\rm L}(k)
	\,.
\label{eq:linearpower}
}
We can also calculate the bispectrum of the redshift-space galaxy as
\al{
\ave{\delta_{{\rm gal},s}(a,{\bm k}_1)\delta_{{\rm gal},s}(a,{\bm k}_2) \delta_{{\rm gal},s}(a,{\bm k}_3)} = (2 \pi)^3 B_s ({\bm k}_1, {\bm k}_2, {\bm k}_3;a)\,\delta^3_{\rm D}({\bm k}_1+{\bm k}_2+{\bm k}_3)
	\,,
}
with
\al{	
	B_s({\bm k}_1,{\bm k}_2,{\bm k}_3;a)
		=2D_+^4(a) Z_1({\bm k}_1;a)Z_1({\bm k}_2;a)Z_2({\bm k}_1,{\bm k}_2;a)P_{\rm L}(k_1)P_{\rm L}(k_2) +({\rm cyc})
	\,.
	\label{eq:rsd_bispectrum}
}

Measuring the growth history of cosmological structures 
can elucidate the nature of the dark energy and
test gravity theory on cosmological scales,
since the modification of gravity theory typically alters the clustering property of nonlinear structure and
the peculiar velocity field.
However, there are a wide variety of gravity theories that yield different signatures to the large-scale structure. 
As shown in Eqs.~\eqref{eq:linearpower} and \eqref{eq:rsd_bispectrum},
the power spectrum and bispectrum is sensitive to the growth of the
structure formation through the RSD kernels.
Hence,
in this paper, we only focus on the general scalar-tensor theory with second-order equations
of motion, namely the Horndeski theory,
and demonstrate the potential of the redshift-space galaxy bispectrum
to constraint the Horndeski theory by future galaxy surveys.

%%%%%%%%%%%%%%%%%%%%%%%%%%%%%%%%%%%%%%%%%%%%%%%%%%%%%%%%%%%%%%%%%%%%%%%%%
%=======================================================================%
\section{Non-linear gravitational growth in Horndeski theory}
\label{sec:Horndeski theory of gravity}
%=======================================================================%
%%%%%%%%%%%%%%%%%%%%%%%%%%%%%%%%%%%%%%%%%%%%%%%%%%%%%%%%%%%%%%%%%%%%%%%%%

The Horndeski theory is a most general scalar-tensor gravity 
theory with the second-order equations of motion, which is paid attention as 
one of attractive modified gravity theories.
Therefore, so far, there are many attempts to test the Horndeski theory
not only in the Solar system but also in the cosmological context.
Up to the second order in cosmological perturbation theory, the
modification of gravity theory would be captured in $F_2$ and $G_2$, which are the second-order kernels appearing in a general formula for the
galaxy bispectrum in redshift space as shown in Eq.~(\ref{eq:sokernel}).
In this section, following a previous work \cite{Takushima:2013foa}, first we briefly review
the derivation of the second-order kernels $F_2$ and $G_2$ in Horndeski theory,
and then we give the expressions by using the effective-field-theory~(EFT) parameters~\cite{Bellini:2014fua} which make us investigate
the deviation from the standard $\Lambda$CDM model with GR easier.

The action of
Horndeski theory of gravity is given by~\cite{Horndeski:1974wa}
\al{
	S=\int\dd^4 x\sqrt{-g}\sum_{a=2}^5{\cal L}_a[g_{\mu\nu},\phi] +S_{\rm m}[g_{\mu\nu},\psi_{\rm m}]
	\,,
\label{eq:lagragian_HD}
}
where the four Lagrangian ${\cal L}_a$ encode the dynamics of the metric $g_{\mu\nu}$ and the scalar field $\phi$. 
These explicit forms are described by
\al{
	&{\cal L}_2=K(\phi ,X)
	\,,\\
	&{\cal L}_3=-G_3(\phi ,X)\Box\phi
	\,,\\
	&{\cal L}_4=G_4(\phi ,X)R+G_{4X}(\phi ,X)\Bigl[(\Box\phi)^2-(\nabla_\mu\nabla_\nu\phi )^2\Bigr]
	\,,\\
	&{\cal L}_5=G_5(\phi ,X)G_{\mu\nu}\nabla^\mu\nabla^\nu\phi
		-\frac{1}{6}G_{5X}(\phi ,X)\Big[(\Box\phi)^3-3\Box\phi(\nabla_\mu\nabla_\nu\phi)^2+2(\nabla_\mu\nabla_\nu\phi)^3\Bigr]
	\,,
}
with $K$ and $G_a~ (a=3,4,5)$ being an arbitrary function of $\phi$ and
$X=-g^{\mu\nu}\pd_\mu\phi\pd_\nu\phi /2$ and
$G_{aX}={\pd G_a/\pd X}$.
In Eq.~\eqref{eq:lagragian_HD}, $S_{\rm m}$
describes the matter sector
and we assume that matter is universally minimal-coupled to the metric and does not
have direct coupling with the scalar field $\phi$.

The Friedmann and matter conservation equations in the Friedmann-Lema\^itre-Robertson-Walker~(FLRW) cosmological background
can be written in the standard manner~\cite{Bellini:2014fua}:
\al{
	&H^2=\frac{1}{3M^2}\left(\rho_{\rm m}+\rho_{\rm DE}\right)
	\,,\\
	&\dot\rho_{\rm m}+3H\rho_{\rm m}=0
	\,,
}
with
\al{
	M^2=2\left( G_4-2XG_{4X}+XG_{5\phi}-\dot\phi HXG_{5X}\right)
	\,,
}
and
\al{
\rho_{\rm DE} =& \,2 X K_{X} - K - 2 X G_{3 \phi} + 6 H \dot{\phi} \left( X G_{3X} - G_{4 \phi}  \right) \nonumber\\
& + 12 H^2 X \left( G_{4 X} + 2 X G_{4 XX} - G_{5 \phi} - X G_{5 \phi X} \right) 
- 4 X H \dot{\phi} \left( 3 G_{4 \phi X} - H^2 \left( G_{5 X} + X G_{5 XX} \right) \right)\,,
}
where the scale $M$ is consider to be about Planck mass scale, and $\rho_{\rm m}$ is the background energy density of matter. 

Let us derive the evolutional 
equations governing the density perturbations in the
flat FLRW with Horndeski theory.
Throughout this paper, we work in the Newtonian gauge which is defined as
\al{
	\dd s^2=-\left( 1+2\Phi (t,{\bm x})\right)\dd t^2+a^2(t)\left( 1-2\Psi (t,{\bm x})\right)\dd{\bm x}^2
	\,,
\label{eq:FLRW}
	}
and the perturbation of the scalar field is described by
\al{
\phi(t,{\bm x}) \to \phi (t) + \delta \phi (t,{\bm x})~.
}
We are interested in  
the behavior of the gravitational and scalar fields on subhorizon scales sourced by
a nonrelativistic matter overdensity.
Therefore, in the derivation of the basic equations, we ignore time derivatives in the effective equations, while keeping spatial
derivatives.
We will keep the nonlinear term schematically written as $(\nabla^2\epsilon )^n$, where $\nabla$ and $\epsilon$
stand for the spatial derivatives and any of $\Phi$\,, $\Psi$\,, and the
perturbation of the scalar field~$\delta\phi$, respectively.
Under these assumptions, the basic equations up to the second-order
({\it i.e.}, $n=1,2$) perturbations are given by~\cite{Kimura:2011dc, Takushima:2013foa}
\al{
	&\nabla^2\left({\cal F}_{\rm T}\Psi -{\cal G}_{\rm T} \Phi -A_1Q\right)
		=\frac{B_1}{2a^2H^2}{\cal Q}^{(2)}
		+\frac{B_3}{a^2H^2}\left(\nabla^2\Phi\nabla^2Q-\nabla_i\nabla_j\Phi\nabla^i\nabla^jQ\right)
	\,,\label{eq:eq1}\\
	&\nabla^2\left({\cal G}_T\Psi +A_2Q\right) =\frac{a^2}{2}\rho_{\rm m}\delta
		-\frac{B_2}{2a^2H^2}{\cal Q}^{(2)}
		-\frac{B_3}{a^2H^2}\left(\nabla^2\Psi\nabla^2Q-\nabla_i\nabla_j\Psi\nabla^i\nabla^jQ\right)
	\,,\label{eq:eq2}\\
	&\nabla^2\left(A_0Q-A_1\Psi-A_2\Phi\right)
		=-\frac{B_0}{a^2H^2}{\cal Q}^{(2)}+\frac{B_1}{a^2H^2}\left(\nabla^2\Psi\nabla^2Q-\nabla_i\nabla_j\Psi\nabla^i\nabla^jQ\right)
	\notag\\
	&\quad\quad\quad
			+\frac{B_2}{a^2H^2}\left(\nabla^2\Phi\nabla^2Q-\nabla_i\nabla_j\Phi\nabla^i\nabla^jQ\right)
			+\frac{B_3}{a^2H^2}\left(\nabla^2\Phi\nabla^2\Psi -\nabla_i\nabla_j\Phi\nabla^i\nabla^j\Psi\right)
	\,,\label{eq:eq3}
}
where we have parameterized the scalar field perturbations as $Q\equiv H\delta\phi /\dot\phi$
and ${\cal Q}^{(2)}=\left(\nabla^2 Q\right)^2-\left(\nabla_i\nabla_j Q\right)^2$.
The coefficients ${\cal G}_{\rm T}$\,, ${\cal F}_{\rm T}$\,, $A_i$\,, and $B_i$ are written in terms of the Horndeski functions $K$ and $G_a$.
The explicit forms of these coefficients are summarized in Appendix~\ref{sec:coefficients}.
In the standard $\Lambda$CDM model with GR, ${\cal F}_T$ and ${\cal G}_T$ can be reduced to $M_{\rm pl}^2$ and all of $A_i$ and $B_i$ become zero,
and hence Eq. (\ref{eq:eq1}) and Eq. (\ref{eq:eq2}) respectively give $\Psi = \Phi$ and the standard Poisson equation.
From the above three equations, the Fourier components of the metric perturbations and the scalar field perturbation, $\Phi$, $\Psi$, and $Q$, can be formally written
as~\cite{Koyama:2009me,Bose:2016qun}
\al{
	-\frac{k^2}{a^2H^2}\epsilon (t,{\bm k})=\kappa_\epsilon (t,k)\delta (t,{\bm k})
		+\int\frac{\dd^3{\bm k}_1\dd^3{\bm k}_2}{(2\pi )^3}
		\delta^3_{\rm D}({\bm k}_1+{\bm k}_2-{\bm k})\gamma_{2,\epsilon} ({\bm k}_1,{\bm k}_2;t)\delta (t,{\bm k}_1)\delta (t,{\bm k}_2)
		+\cdots
	\,,\label{eq:Phi assumption}
}
up to the second order in the matter density perturbations.
Each time dependent coefficient in the linear term, $\kappa_\epsilon$, can be expressed as
\al{
	&\kappa_\Phi (t)=\frac{\rho_{\rm m}(A_0{\cal F}_{\rm T}-A_1^2)}{2H^2(A_0{\cal G}_{\rm T}^2+2A_1A_2{\cal G}_{\rm T}+A_2^2{\cal F}_{\rm T})}
	\,,\label{eq:kappa_Phi}\\
	&\kappa_\Psi (t)=\frac{\rho_{\rm m}(A_0{\cal G}_{\rm T}-A_1A_2)}{2H^2(A_0{\cal G}_{\rm T}^2+2A_1A_2{\cal G}_{\rm T}+A_2^2{\cal F}_{\rm T})}
	\,,\\
	&\kappa_Q (t)=\frac{\rho_{\rm m}(A_1{\cal G}_{\rm T}-A_2{\cal F}_{\rm T})}{2H^2(A_0{\cal G}_{\rm T}^2+2A_1A_2{\cal G}_{\rm T}+A_2^2{\cal F}_{\rm T})}
	\,.\label{eq:kappa_Q}
}
Moreover,
comparing the second-order contributions of Eqs.~\eqref{eq:eq1}-\eqref{eq:eq3} with Eq.~\eqref{eq:Phi assumption},
we obtain the nonlinear interaction term  for $\Phi$ as %, {\it e.g.},
\al{
	\gamma_{2,\Phi} ({\bm k}_1,{\bm k}_2;t)=\tau_\Phi (t)	\left(
	1-\frac{({\bm k}_1\cdot{\bm k}_2)^2}{k_1^2k_2^2} \right)
	%\gamma ({\bm k}_1,{\bm k}_2)
	\,,
\label{eq:gamma2_Phi} %\\
}
where the time-dependent coefficient $\tau_\Phi$ is %we have defined
\al{
	\tau_\Phi (t)=\frac{H^2}{\rho_{\rm m}}\left( 2B_0\kappa_Q^3-3B_1\kappa_\Psi\kappa_Q^2-3B_2\kappa_\Phi\kappa_Q^2-6B_3\kappa_\Phi\kappa_\Psi\kappa_Q\right)
	\,.\label{eq:tau_Phi} %\\
%	&\gamma ({\bm k}_1,{\bm k}_2)
%		=\alpha^{\rm sym}({\bm k}_1,{\bm k}_2)-\beta ({\bm k}_1,{\bm k}_2)
%		=1-\frac{({\bm k}_1\cdot{\bm k}_2)^2}{k_1^2k_2^2}
%	\,.
}
In the similar way, we can obtain $\kappa$ and $\tau$ for $\Psi$ and
$Q$.
However,
since we consider matter which is
minimal-coupled to the metric,
the effect of the gravity theory
on the evolution of the matter perturbations appears only through the
gravitational potential $\Phi$ as same as in the GR case.
%%%%%%%%%%%%% reduced by SY %%%%%%%%%%%%% 
% shown
%in~Eqs.~\eqref{eq:delta eq}~and~\eqref{eq:theta eq}.
%%%%%%%%%%%%%%%%%%%%%%%%%%%%%%%%%%%%%%%%%
Therefore, only $\kappa$ and $\tau$ for $\Phi$ are required
to evaluate the evolution of the matter perturbation in the second-order
perturbation theory.
Note that,
although these coefficients, $\kappa$ and $\tau$, should be in general treated as the scale-dependent terms when the scalar potential terms are taken into account 
to accommodate chameleon-type models such as $f(R)$ gravity~\cite{Bose:2016qun}, we only consider the Vainshtein-type models in which $\kappa$ and $\tau$
are shown to depend only on the time.

Next we consider the matter perturbation evolution.
In the case of minimal-coupling,
the evolution equations for the matter perturbations are same as in the
case of GR. 
Then,
the modification of the gravity theory in the growth of the matter density fluctuations
would appear through $\kappa_\Phi$ and $\gamma_{2, \Phi} $ $({\rm or}~\tau_\Phi)$ given in the above discussion. Here, we focus on the
galaxy bispectrum which can probe the nonlinear (second-order) gravitational evolution of the matter density fluctuations,
and it is described by
the second-order perturbative kernels $F_2$ and $G_2$
defined in Eqs.~\eqref{eq:def_kernel_F} and~\eqref{eq:def_kernel_G}.
It has been found that in Horndeski theory these kernels
are provided in the more suggestive forms as~\cite{Takushima:2013foa,Takushima:2015iha}
\al{
	&F_2({\bm k}_1,{\bm k}_2)=\left(1+\frac{{\bm k}_1\cdot{\bm k}_2(k_1^2+k_2^2)}{2k_1^2k_2^2}\right) -\frac{2}{7}\lambda (t)\left(1-\frac{({\bm k}_1\cdot{\bm k}_2)^2}{k_1^2k_2^2} \right)
	\,,\\
	&G_2({\bm k}_1,{\bm k}_2)=\left(1+\frac{{\bm k}_1\cdot{\bm k}_2(k_1^2+k_2^2)}{2k_1^2k_2^2}\right)-\frac{4}{7}\lambda_\theta (t)\left(1-\frac{({\bm k}_1\cdot{\bm k}_2)^2}{k_1^2k_2^2} \right)
	\,,
}
where 
\al{
\lambda_\theta (t)=\lambda (t)+\frac{\dot\lambda (t)}{2f(t)H(t)}	\,.
}
Here, the time-dependent coefficient $\lambda (t)$ obeys a following second-rank differential equation \cite{Takushima:2015iha} (see also \cite{Takushima:2013foa}):
\al{
	\frac{\dd^2\lambda}{\dd\ln a^2}+\left( 2+\frac{\dd\ln H}{\dd\ln a}+4f\right)\frac{\dd\lambda}{\dd\ln a}
		+\left( 2f^2+\kappa_\Phi\right)\lambda
		=\frac{7}{2}\left( f^2-\tau_\Phi\right)
	\,.\label{eq:lambda eq}
}
One can clearly see that $\lambda$ can be induced by $\kappa_\Phi$ and $\tau_\Phi$.
In the above equation, $f$ is the linear growth rate and the evolution equation for $f$ is given by
\al{
	\frac{\dd f}{\dd\ln a}+\left( 2+\frac{\dd\ln H}{\dd\ln a}\right) f+f^2-\kappa_\Phi=0
	\,.\label{eq:f eq}
}
This means that
the precise measurement of the linear growth rate,~$f$,
can test 
the information captured in $\kappa_\Phi$.
On the other hand, a new coefficient $\lambda$ appearing in the
second-order kernels depends on not only $\kappa_\Phi$ but
also $\tau_\Phi$ which could have new information about the modification of gravity theory. Thus, in this sense, we would like to stress
that the precise investigation of the nonlinear growth of matter density fluctuations by using galaxy bispectrum
would give new insight into the gravity theory.
Although $\lambda$ is shown to go to unity in the limit of the Einstein-de Sitter Universe,
the large deviation from unity can be possible in the Horndeski theory, as seen in the subsequent section. 
Once the Horndeski functions $G_a$ are given as the underlying model, we can solve the equations for the growth rate and
the coefficient of the second-order kernel by substituting Eqs.~\eqref{eq:kappa_Phi}-\eqref{eq:kappa_Q} and \eqref{eq:tau_Phi} 
into Eqs.~\eqref{eq:lambda eq} and \eqref{eq:f eq}.

%%%%%%%%%%%%%%%%%%%%%%%%%%%%%%%%%%%%%%%%%%%%%%%%%%%%%%%%%%%%%%%%%%%%%%%%%
%=======================================================================%
\section{Analytic evaluation}
\label{sec:analytic evaluation}
%=======================================================================%
%%%%%%%%%%%%%%%%%%%%%%%%%%%%%%%%%%%%%%%%%%%%%%%%%%%%%%%%%%%%%%%%%%%%%%%%%

As shown in the previous section,  the growth rate $f$ and the coefficient of the second-order kernel $\lambda$
would be good candidates to capture the signature of the modification of gravity theory.
The general solutions of $f$ and $\lambda$ are formal and the attempting
to fit observations of the growth history 
for each model is rather complicated.
One alternatively considers the simple characterization stimulating physical intuition.

%%%%%%%%%%%%%%%%%%%%%%%%%%%%%%%%%%%%%%%%%%%%%%%%%%%%%%%%%%%%%%%%%%%%%%%%%%%%%%
%=======================================================================%
\subsection{EFT parametrization of Horndeski gravity}
\label{sec:Effecitve-theory approach}
%=======================================================================%
%%%%%%%%%%%%%%%%%%%%%%%%%%%%%%%%%%%%%%%%%%%%%%%%%%%%%%%%%%%%%%%%%%%%%%%%%
First,
instead of considering the explicit form of the Horndeski function $G_a$, we will introduce the 
EFT parameters
to specify the cosmological information.
In Ref.~\cite{Bellini:2014fua}, the authors identified the minimum number of the functions that fully specify these linear perturbations 
in the Horndeski class of gravity theory.
The minimum set to specify the total amount of cosmological information
up to the linear order is 
four functions of time 
that can be labeled $\{\alpha_{\rm M}\,,\alpha_{\rm T}\,,\alpha_{\rm B}\,,\alpha_{\rm K}\}$
in addition to the background expansion history $H(t)$ and the effective Planck mass $M(t)$.
These $\alpha_i$ are related to the Horndeski variables as~\cite{Bellini:2014fua,Ade:2015rim}
\al{
	&HM^2\alpha_{\rm M}=\frac{\dd}{\dd t}M^2
	\,,\label{eq:alpha_M}\\
	&M^2\alpha_{\rm T}=2X\biggl[ 2G_{4X}-2G_{5\phi}-\left(\ddot\phi -H\dot\phi\right) G_{5X}\biggr]
	\,,\\
	&HM^2\alpha_{\rm B}=2\dot\phi\left( XG_{3X}-G_{4\phi}-2XG_{4\phi X}\right)+8HX\left( G_{4X}+2XG_{4XX}-G_{5\phi}-XG_{5\phi X}\right)
	\notag\\
	&\quad\quad\quad\quad\quad
		+2\dot\phi H^2X\left( 3G_{5X}+2XG_{5XX}\right)
	\,,\\
	&H^2M^2\alpha_{\rm K}
		=2X\left( K_X+2XK_{XX}-2G_{3\phi}-2XG_{3\phi X}\right)
			+12\dot\phi HX\left(G_{3X}+XG_{3XX}-3G_{4\phi X}-2XG_{4\phi XX}\right)
	\notag\\
	&\quad\quad\quad\quad\quad
			+12H^2X\left(G_{4X}+8XG_{4XX}+4X^2G_{4XXX}-G_{5\phi}-5XG_{5\phi X}-2X^2G_{5\phi XX}\right)
	\notag\\
	&\quad\quad\quad\quad\quad
			+4\dot\phi H^3X\left( 3G_{5X}+7XG_{5XX}+2X^2G_{5XXX}\right)
	\,.
}
The $\alpha$ functions represent the linear freedom of the Horndeski class of models.
Moreover, to generalize this fact to the second-order perturbation theory in the subhorizon scales, 
two other functions of time are required~\cite{Bellini:2015wfa}.
To fit the second-order perturbations, we further define new two $\alpha$ functions as
\al{
	&M^2\alpha_{\rm V1}=-2X\left( G_{4X}+2XG_{4XX}-G_{4\phi}+2H\dot\phi G_{5X}-XG_{5\phi X}+\dot\phi HXG_{5XX}\right)
	\,,\\
	&M^2\alpha_{\rm V2}=2\dot\phi HXG_{5X}
	\,,\label{eq:alpha_V2}
}
where $\alpha_{\rm V1}$ and $\alpha_{\rm V2}$ represent the amplitude of the Vainshtein screening 
mechanism~\cite{Kimura:2011dc,Narikawa:2013pjr,Koyama:2013paa,Kobayashi:2014ida}.  
To derive the effective theory in terms of the scalar field perturbations, we simply consider
the form of the scalar field:
\al{
	\phi (t)\to\phi =t+\delta\phi (t,{\bm x})
	\,.
}
Using these $\alpha$ functions, the coefficients in the basic equations \eqref{eq:eq1}-\eqref{eq:eq3} can be rewritten as
\al{
	&{\cal G}_{\rm T}=M^2
	\,,\\
	&{\cal F}_{\rm T}=M^2\left( 1+\alpha_{\rm T}\right)
	\,,\\
	&A_0=M^2\biggl[\frac{\dot H}{H^2}+\frac{3}{2}\widetilde\Omega_{\rm m}-\frac{(HM^2\alpha_{\rm B})^\cdot}{2H^2M^2}-\left(\alpha_{\rm M}-\alpha_{\rm T}+\frac{1}{2}\alpha_{\rm B}\right)\biggr]
	\,,\\
	&A_1=M^2\left(\alpha_{\rm M}-\alpha_{\rm T}\right)
	\,,\\
	&A_2=\frac{1}{2}M^2\alpha_{\rm B}
	\,,\\
	&B_0=\frac{1}2{}M^2\biggl[
				\alpha_{\rm V1}-\frac{(M^2\alpha_{\rm V1})^\cdot}{HM^2}
				+\alpha_{\rm M}+\alpha_{\rm B}
				-\frac{3}{2}\biggl\{\alpha_{\rm T}-\alpha_{\rm V2}+\frac{1}{M^2}\left(\frac{M^2\alpha_{\rm V2}}{H}\right)^\cdot\biggr\}
				\biggr]
	\,,\\
	&B_1=\frac{1}{2}M^2\Biggl[\alpha_{\rm T}-\alpha_{\rm V2}+\frac{1}{M^2}\left(\frac{M^2\alpha_{\rm V2}}{H}\right)^\cdot\Biggr]
	\,,\\
	&B_2=M^2\left(\alpha_{\rm V1}+\alpha_{\rm V2}\right)
	\,,\\
	&B_3=\frac{1}{2}M^2\alpha_{\rm V2}
	\,,
}
where we have introduced the fractional matter density defined as
\al{
	\widetilde\Omega_{\rm m}=\frac{\rho_{\rm m}}{3M^2H^2}
	\,.
}
As pointed out in Ref.~\cite{Bellini:2014fua}, the sound speed of the scalar field perturbations is sufficiently close to the speed of light, 
the quasi-static approximation is valid and $\alpha_{\rm K}$ does not appear the equations for the matter density fluctuations.

%%%%%%%%%%%%%%%%%%%%%%%%%%%%%%%%%%%%%%%%%%%%%%%%%%%%%%%%%%%%%%%%%%%%%%%%%%%%%%
%=======================================================================%
\subsection{Approximate expressions}
\label{sec:approximate expressions}
%=======================================================================%
%%%%%%%%%%%%%%%%%%%%%%%%%%%%%%%%%%%%%%%%%%%%%%%%%%%%%%%%%%%%%%%%%%%%%%%%%

Next, we introduce phenomenological parameters that describe non-standard cosmological models.
For the growth rate $f$, the gravitational growth index, $\gamma$,  is known as the simplest parametrization:
\al{
	f(a)=\widetilde\Omega_{\rm m}^\gamma (a)
	\,.
}
One can capture the signature of the deviations from the general relativity up to the linear order through the difference in $\gamma$
from the standard value $\gamma_{\rm GR+\Lambda CDM}\approx 6/11$.
Let us derive the analytic formula of the gravitational growth index for the Horndeski class of gravity theory~\cite{Linder:2007hg}.
Substituting this parametrization for $f(a)$ into Eq.~\eqref{eq:f eq} yields
\al{
	\gamma\,\widetilde\Omega_{\rm m}^{\gamma}\frac{\dd\ln\widetilde\Omega_{\rm m}}{\dd\ln a}
		+\left( 2+\frac{\dd\ln H}{\dd\ln a}\right)\widetilde\Omega_{\rm m}^\gamma +\widetilde\Omega_{\rm m}^{2\gamma}-\kappa_\Phi =0
	\,.\label{eq:gamma eq}
}
To evaluate the value of $\gamma$, it is required to specify the expansion history of the Universe.
Therefore we rewrite the Friedmann and matter conservation equations in terms of $\widetilde\Omega_{\rm m}$ as
\al{
	&\frac{\rho_{\rm DE}}{3M^2M^2}=1-\widetilde\Omega_{\rm m}
	\,,\\
	&\frac{\dd\ln H}{\dd\ln a}
		=-\frac{3}{2}\biggl[ 1+w_{\rm DE}\left( 1-\widetilde\Omega_{\rm m}\right)\biggr]
	\,,\label{eq:dLogHdLoga eq}\\
	&\frac{\dd\ln\widetilde\Omega_{\rm m}}{\dd\ln a}
		=3w_{\rm DE}\left( 1-\widetilde\Omega_{\rm m}\right) -\alpha_{\rm M}
	\,,\label{eq:dLogOmegamdLoga eq}
}
where $w_{\rm DE}$ is an effective equation of state for the dark energy component, which is defined as $w_{\rm DE} := p_{\rm DE}/\rho_{\rm DE}$
with $p_{\rm DE} := - M^2 (3 H^2 + 2 \dot{H})$.
We assume that the Universe can be well described by the $\Lambda$CDM model in the deep matter dominated era,
hence $w_{\rm DE}$ and $\kappa_\Phi$ can be approximated in the following expanded forms:
\al{
	&w_{\rm DE}=\sum_{n=0}\frac{1}{n!}w^{(n)}\left( 1-\widetilde\Omega_{\rm m}\right)^n
	\,,\\
	&\kappa_\Phi -\frac{3}{2}\widetilde\Omega_{\rm m} =\sum_{n=1}^\infty\frac{1}{n!}\kappa_\Phi^{(n)}\left( 1-\widetilde\Omega_{\rm m}\right)^n
	\,.\label{eq:kappa_Phi exp}
}
For $\alpha_{\rm M}$, we take the following parametrization as
\al{
	&\alpha_{\rm M}=c_{\rm M}\left( 1-\widetilde\Omega_{\rm m}\right)
	\,.\label{eq:alpha_M def}
}
Combining Eqs.~\eqref{eq:gamma eq}-\eqref{eq:alpha_M def}, the leading order equation in the limit of $1-\widetilde\Omega_{\rm m}\ll 1$ leads to 
the approximate value of the gravitational growth index
\al{
	\gamma\approx \frac{3-3w^{(0)}-2\kappa_\Phi^{(1)}}{5-6w^{(0)}+2c_{\rm M}}
	\,.\label{eq:gamma estimation}
}
This result is a generalization of the well-known formula for the $w$CDM model~\cite{Linder:2007hg}.
We note that in the case of the $\Lambda$CDM model with GR as our fiducial model, 
we take $w^{(0)}=-1$\,, $\kappa_\Phi^{(1)}=c_{\rm M}=0$ and the standard
result $\gamma =6/11\approx 0.545$ can be recovered. 
Current observations give the stringent constraint on $\gamma$. 
In particular, the deviation from the standard value is allowed only below $10\%$.
Hence, the only small deviations of $(w^{(0)}+1)$, $c_{\rm M}$ and $\kappa_\Phi^{(1)}$ are possible.

Similarly, for the coefficient of the second-order kernel, $\lambda$, we
propose the following approximated form as the simplest characterization
with the second-order index~$\GG$:
\al{
	\lambda (t)=\widetilde\Omega_{\rm m}^\GG (t)
	\,.\label{eq:lambda-Gamma def}
}
Applying this, we rewrite Eq.~\eqref{eq:lambda eq} to
\al{
	\widetilde\Omega_{\rm m}^\GG
		\Biggl[
			\left(\GG\frac{\dd\ln\widetilde\Omega_{\rm m}}{\dd\ln a}\right)^2
			+\GG\frac{\dd^2\ln\widetilde\Omega_{\rm m}}{\dd\ln a^2}
			+\left( 2+\frac{\dd\ln H}{\dd\ln a}+4\widetilde\Omega_{\rm m}^{4\gamma}\right)\GG\frac{\dd\ln\widetilde\Omega_{\rm m}}{\dd\ln a}
			+\left( 2\widetilde\Omega_{\rm m}+\kappa_\Phi\right)
		\Biggr]
		=\frac{7}{2}\left(\widetilde\Omega_{\rm m}^{2\gamma}-\tau_\Phi\right)
	\,.\label{eq:lambda eq2}
}
Since $\tau_\Phi$ reduces to zero in the limit of the $\Lambda$CDM+GR, we can expand 
\al{
	&\tau_\Phi =\sum_{n=1}^\infty\frac{1}{n!}\tau_\Phi^{(n)}\left( 1-\widetilde\Omega_{\rm m}\right)^n
	\,.\label{eq:tau_Phi exp}
}
Substituting this expression and Eqs.~\eqref{eq:dLogHdLoga eq}-\eqref{eq:alpha_M def} into Eq.~\eqref{eq:lambda eq2}, we find that 
the leading value of the second-order index 
can be estimated as
\al{
	\GG
		=\frac{-3+6\gamma+2\kappa_\Phi^{(1)}+7\tau_\Phi^{(1)}}{(7-6w^{(0)}+2c_{\rm M})(1-3w^{(0)}+c_{\rm M})}
	\,.\label{eq:Gamma estimation}
}
In the case of the $\Lambda$CDM model with GR, $\tau_\Phi^{(1)}=0$ and we have $\GG =3/572\approx 0.00524$, 
suggesting that $\lambda$ can be well approximated by unity in this limit~\cite{Scoccimarro:2000ee}.
Once the underlying theory describing the dark energy or the modification of gravity theory is given,
we immediately calculate the second-order index~$\GG$ through Eq.~\eqref{eq:Gamma estimation}
as well as the gravitational growth index $\gamma$, Eq.~\eqref{eq:gamma estimation}.

In the above discussion, we have obtained approximated forms of $\gamma$ and $\xi$ which capture the modification of gravity theory, in terms of
$w_{\rm DE}$, $\kappa_\Phi$, $\tau_\Phi$ and an EFT function $\alpha_{\rm M}$. Since $\kappa_\Phi$ and $\tau_\Phi$ are written in terms of
the EFT $\alpha$ functions introduced in the previous subsection, the phenomenological parameters $\gamma$ and $\xi$ are approximately written with
the EFT functions.
We take a following parametrization for all $\alpha$ functions, which is suggested in Ref.~\cite{Bellini:2014fua}:
\al{
	\alpha_i =c_i\left( 1-\widetilde\Omega_{\rm m}\right)
	\,,\label{eq:alpha func}
}
where $c_i$ are constant characterizing the amplitude of the modification.
Hence the remaining nontrivial parameters can be reduced to $\{c_{\rm B}\,,c_{\rm M}\,,c_{\rm T}\,,c_{\rm V1}\,,c_{\rm V2}\}$.

\begin{figure}[t]
\includegraphics[width=100mm]{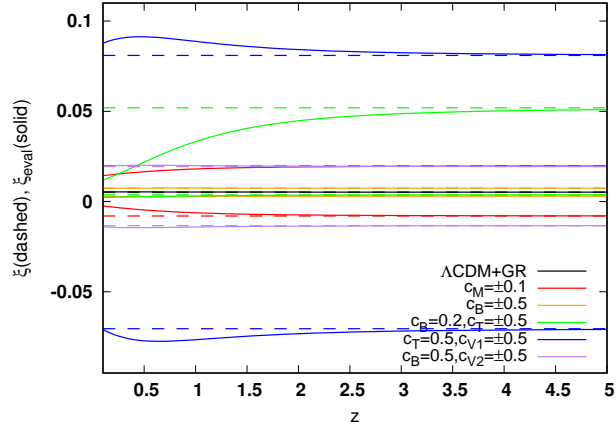}
\caption{
The evolution of $\GG$ for different $\alpha$ functions.
For comparison, we also plot the estimated value of $\GG$ through
Eq.~\eqref{eq:Gamma estimation}. In most of cases, $\GG$ is independent on the redshift.
}
\label{fig:Gamma-z}
\end{figure}

Using the concrete expression for the~$\alpha$~functions in Eq.~\eqref{eq:alpha func},
we can apply the formalism derived in the previous subsection to calculate $\lambda$ and $\GG$.
Therefore, we first explicitly expand the quantities in terms of $1-\widetilde\Omega_{\rm m}$ as
\al{
	&\kappa_\Psi -\frac{3}{2}\widetilde\Omega_{\rm m}
		=\sum_{n=1}^\infty\frac{1}{n!}\kappa_\Psi^{(n)}\left( 1-\widetilde\Omega_{\rm m}\right)^n
	\,,\\
	&\kappa_Q
		=\sum_{n=0}^\infty\frac{1}{n!}\kappa_Q^{(n)}\left( 1-\widetilde\Omega_{\rm m}\right)^n\,.
}
By using these constant parameters, we then obtain the leading-order corrections:
\al{
	&\kappa_\Phi^{(1)}
		=\frac{3}{2}
			\biggl[
			c_{\rm T}
			+\frac{(c_{\rm B}+2c_{\rm M}-2c_{\rm T})^2}{6(1+w^{(0)})-6c_{\rm B}w^{(0)}+2c_{\rm B}c_{\rm M}-c_{\rm B}+4c_{\rm M}-4c_{\rm T}}
			\biggr]
	\,,\label{eq:kappa_Phi^1}\\
	&\kappa_Q^{(0)}
		=-\frac{3(c_{\rm B}+2c_{\rm M}-2c_{\rm T})}
		{6(1+w^{(0)})-6c_{\rm B}w^{(0)}+2c_{\rm B}c_{\rm M}-c_{\rm B}+4c_{\rm M}-4c_{\rm T}}
	\,,\\
	&\kappa_\Psi^{(1)}
		=-\frac{c_{\rm B}}{2}\kappa_Q^{(0)}
	\,,
}
and
\al{
	\tau_\Phi^{(1)}=&\frac{1}{3}\left(\kappa_Q^{(0)}\right)^2\biggl[\left(c_{\rm B}+c_{\rm M}-\frac{3}{2}c_{\rm T}\right)\kappa_Q^{(0)}-\frac{9}{4}c_T\biggr]
	\notag\\
	&
					+\frac{1}{3}\left(\kappa_Q^{(0)}\right)^2\biggl[\left( 1+3w^{(0)}-c_{\rm M}\right)\kappa_Q^{(0)}-\frac{9}{2}\biggr]c_{\rm V1}
	\notag\\
	&
					-\frac{1}{2}\kappa_Q^{(0)}\left(\kappa_Q^{(0)}+\frac{3}{2}\right)\biggl[\left(\frac{1}{2}-3w^{(0)}+c_{\rm M}\right)\kappa_Q^{(0)}+3\biggr]c_{\rm V2}
	\,.\label{eq:tau_Phi^1}
}
By substituting these into 
Eqs.~\eqref{eq:gamma estimation} and \eqref{eq:Gamma estimation},
we can evaluate the indices of the gravitational growth and the second-order kernel, $\gamma$ and $\xi$,
in terms of constant parameters $\{ w^{(0)}\,, c_{\rm B}\,,c_{\rm M}\,,c_{\rm T}\,,c_{\rm V1}\,,c_{\rm V2}\}$,
  while the explicit expressions are complicated.
To confirm the validity of the approximated expression for our new parameter $\xi$, 
at least, we have to check
the constancy of $\GG$. 
In Fig.~\ref{fig:Gamma-z},
we numerically evaluate $\GG_{\rm eval}(t)=\ln\lambda (t)/\ln\widetilde\Omega_{\rm m}(t)$
with various values of the $\alpha$ functions.
We found that the deviations from the constant value of $\GG_{\rm eval}$ in most cases are less than $10\%$, though in some cases
the constancy of $\GG$ seems to be eventually violated in low redshifts.
Hence we conclude that we can neglect the time-dependence on $\GG$
as far as future surveys observing high redshifts are considered.

%>>>>>>>>>>>>>>>>>>>>>>>>>>>>>>>FIGURE<<<<<<<<<<<<<<<<<<<<<<<<<<<<<<<<<<%
%>>>>>>>>>>>>>>>>>>>>>>>>>>>>>>>>>><<<<<<<<<<<<<<<<<<<<<<<<<<<<<<<<<<<<<%

Let us consider a specific cosmological model to investigate
the gravitational growth index~$\gamma$ and the second-order index~$\GG$.
We assume the $\Lambda$CDM cosmology as the background Hubble expansion, namely $w^{(0)}=-1$.
In particular, when we take the small braiding limit, namely $c_{\rm B}\to 0$ the situation is found to be drastically simplified (see also Ref.~\cite{Bellini:2015wfa}).
The parameters in Eqs.~\eqref{eq:kappa_Phi^1}-\eqref{eq:tau_Phi^1} reduce to
\al{
	\kappa_\Phi^{(1)}=\frac{3}{2}c_{\rm M}
	\,,\ \ \ 
	\kappa_Q^{(0)}=-\frac{3}{2}
	\,,\ \ \ 	
	\kappa_\Psi^{(1)}=0
	\,,\ \ \ 
	\tau_\Phi^{(1)}=\frac{9}{8}\Bigl[\left( c_{\rm M}-1\right) c_{\rm V1}-c_{\rm M}\Bigr]
	\,.
}
Surprisingly, these variables depend only on $c_{\rm M}$ and $c_{\rm
V1}$.
Hence
$\gamma$ and $\GG$
also depend only on  $c_{\rm M}$ and $c_{\rm V1}$ as
\al{
&\gamma (c_{\rm B}=0)=\frac{3(2-c_{\rm M})}{11+2c_{\rm M}}
\,,\label{eq:gamma0}\\
&\GG (c_{\rm B}=0)
=\frac{3}{8(4+c_{\rm M})(13+2c_{\rm M})}
\biggl[
\frac{(8+c_{\rm M})(1-26c_{\rm M})}{11+2c_{\rm M}}+21\left( c_{\rm M}-1\right) c_{\rm V1}
\biggr]
\,.\label{eq:Gamma0}
}
From the above expression, we can easily find that by taking $c_{\rm M} \to 0$
the growth index $\gamma$ goes to $6/11$ which is just the value in the standard $\Lambda$CDM model with GR.
 On the other hand, our new parameter $\xi$ depends not only on $c_{\rm M}$ but also on $c_{\rm V1}$,
 and hence we could realize the large $\GG$ 
even in the case which can not be distinguished with the standard $\Lambda$CDM model with GR up to the linear growth of density fluctuations.
In Appendix \ref{sec:Large Gamma model}, we construct the explicit model in which there is observably
a large deviation for $\xi$ but $\gamma$ remains the standard one.

%%%%%%%%%%%%%%%%%%%%%%%%%%%%%%%%%%%%%%%%%%%%%%%%%%%%%%%%%%%%%%%%%%%%%%%%%
%=======================================================================%
\section{Fisher analysis}
\label{sec:Fisher analysis}
%=======================================================================%
%%%%%%%%%%%%%%%%%%%%%%%%%%%%%%%%%%%%%%%%%%%%%%%%%%%%%%%%%%%%%%%%%%%%%%%%%

Let us numerically investigate the expected constraints on the
second-order index $\GG$  
as well as the gravitational growth index~$\gamma$, based on the Fisher analysis.
To evaluate the expected constraints from the galaxy bispectrum measured in future galaxy surveys, we calculate the Fisher matrix for the bispectrum, 
which is obtained by summing over all possible triangular configurations.
The explicit expression is written as
\al{
	F_{\alpha\beta}
		=\sum_{k_1,k_2,k_3=k_{\rm min}}^{k_{\rm max}}\frac{\pd{\bm B}^{\rm obs}}{\pd\theta^\alpha}\cdot
			\biggl[{\rm Cov}^{-1}({\bm B}^{\rm obs},{\bm B}^{\rm obs})\biggr]\cdot\frac{\pd{\bm B}^{\rm obs}}{\pd\theta^\beta}
	\,,
}
where  $\theta^\alpha$ are free parameters to be determined by observations.
The marginalized expected $1\sigma$ error on parameter $\theta^\alpha$ from the Fisher matrix
is estimated to be $\sigma (\theta^\alpha )=\sqrt{(F^{-1})_{\alpha\alpha}}$.
Assuming the Gaussian covariance, we obtain the covariance matrix as~\cite{Sefusatti:2007ih,Sefusatti:2006pa,Scoccimarro:2003wn}
\al{
	{\rm Cov}({\bm B}^{\rm obs},{\bm B}^{\rm obs})
		=\frac{s_{\rm B}V_{\rm survey}}{N_{\rm t}}
			\left( P_s({\bm k}_1)+\frac{1}{n_{\rm g}}\right)\left( P_s({\bm k}_2)+\frac{1}{n_{\rm g}}\right)\left( P_s({\bm k}_3)+\frac{1}{n_{\rm g}}\right)
	\,,
}
where $s_{\rm B}$ is the symmetric factor describing the number of a given bispectrum
triangle ($s_{\rm B}=6,2$, and $1$ for equilateral, isosceles and general triangles, respectively)
and the quantity $N_{\rm t}=V_{\rm B}/k_{\rm F}^6$ denotes the total number of available 
triangles with $k_{\rm F}=2\pi /V_{\rm survey}^{1/3}$ and $V_{\rm B}=8\pi^2 k_1k_2k_3k_{\rm F}^3$
being the fundamental frequency and the volume of the fundamental cell in Fourier space, respectively.
The maximal wavelength is chosen so that $k_{\rm max}=\pi /(2R_{\rm min})$ 
with $\sigma (R_{\rm min},z)=0.5$ \cite{Sefusatti:2007ih}.
To parametrize the background evolution, we take the constant dark energy equation-of-state parameter for simplicity.
Throughout this paper, our fiducial model is the
$\Lambda$CDM cosmological model with a
nearly scale-invariant primordial scalar perturbations; 
$\Omega_{\rm m,0}=0.318$\,, $\Omega_{\rm b,0}=0.495$\,, $\Omega_{\Lambda
,0}=0.6817$\,, $w_{\rm DE}=-1$\,,
$h=0.67$\,, $n_{\rm s}=0.9619$\,, 
and $\sigma_8 =0.835$.

%>>>>>>>>>>>>>>>>>>>>>>>>>>>>>>>FIGURE<<<<<<<<<<<<<<<<<<<<<<<<<<<<<<<<<<%
\begin{figure}[t]
\includegraphics[width=100mm]{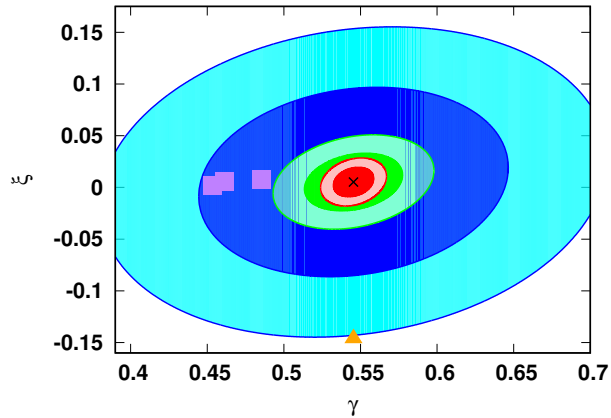}
\caption{
Forecast $1\sigma$ and $2\sigma$ marginalized contours in $(\gamma,\GG )$ plane
for SKA1MID(blue), SKA2(red) and Euclid(green),
marginalizing over the equation-of-state parameter and bias parameters.
For comparison, we also plot the kinetic gravity braiding model (purple boxes),
and the large $\xi$ model (orange triangle).
}
\label{fig:gamma-Gamma_Fisher}
\end{figure} 
%>>>>>>>>>>>>>>>>>>>>>>>>>>>>>>>>>><<<<<<<<<<<<<<<<<<<<<<<<<<<<<<<<<<<<<%

We apply our Fisher matrix analysis to future redshift surveys conducted
by Euclid and the SKA.
Our forecast is performed for the parameter set : $\{ w_{\rm DE}\,, %w_0\,, w_a\,, 
\gamma\,, \GG\,, b_1(z)\,, b_2(z)\}$.
We adopt the predicted number density of galaxies, $n_{\rm g}(z)$, as a
function of redshift, given in Table 3 of  Ref.~\cite{Amendola:2016saw} for Euclid
and in Table 1 of  Ref.~\cite{Bull:2015nra} for the SKA1MID and SKA2, respectively.
Considering the $5$\,, $17$\,, and $14$ redshift bins for SKA1MID, SKA2, and Euclid, in total,
we include $10$\,, $34$\,, $28$ nuisance parameters to model the bias parameters as well as
%four 
three parameters characterizing the growth history.
The fiducial value of $\gamma$ and $\GG$ are $11/6$ and $3/572$, respectively.
The fiducial values of the linear and nonlinear bias parameters are
obtained from the dark matter halo bias,
because galaxies are formed in dark matter halos.
To evaluate galaxy bias parameters, we follow the procedure of the halo bias parameters $b_\ell^{\rm h}(M,z)\ \ (\ell =1,2)$
given in Ref.~\cite{Scoccimarro:2000gm}, that is,
$b_\ell =\frac{1}{n_{\rm g}}\int_{M_{\rm min}}\dd  M\frac{\dd n}{\dd M}b_\ell^{\rm h}\ave{N}_M$\,.
Here we adopt the Sheth-Tormen mass function $\dd n/\dd M$ \cite{Sheth:1999mn}, 
and the halo occupation distributions $\ave{N}_M$ in~Ref.~\cite{Tinker:2004gf}
with the fitting parameters given in Ref.~\cite{Conroy:2005aq},
and we obtain the minimum mass~${M_{\rm min}}$ from
$n_{\rm g}=\int_{M_{\rm min}}\dd M\frac{\dd n}{\dd M}\ave{N}_M$
for a galaxy number density $n_{\rm g}$.

%TTTTTTTTTTTTTTTTTTTTTTTTTTTTTTTTTTTTTTTTTTTTTTTTTTTTTTTTTTTTTTTTTTTTTTTTT%
\begin{table*}
\caption{
Summary of the $1\sigma$ constraints on the equation-of-state
 parameter\,, the gravitational growth index $\gamma$
and the second-order index $\GG$ 
marginalized over the linear and nonlinear bias parameters.}
\vspace{10pt}
 \centering
 \begin{tabular}{cccc}\hline\hline
survey & $\Delta w_{\rm DE}$ & $\Delta \gamma$ & $\Delta \GG$ \\
  \hline
SKA1MID & $0.135$ & $0.067$ & $0.060$ \\
SKA2       & $0.0085$ & $0.0087$ & $0.0094$ \\
Euclid      & $0.016$ & $0.021$ & $0.018$ \\
  \hline\hline
\end{tabular}
\label{table:constnraint}
\end{table*}
%TTTTTTTTTTTTTTTTTTTTTTTTTTTTTTTTTTTTTTTTTTTTTTTTTTTTTTTTTTTTTTTTTTTTTTTTT%

In Fig.~\ref{fig:gamma-Gamma_Fisher}, we show the $1\sigma$ and $2\sigma$ confidence regions
of the gravitational growth index $\gamma$ and the second-order index $\GG$.
The results of our Fisher analysis marginalizing over the bias parameters are summarized in 
Table \ref{table:constnraint}.
For comparison, we also plot the predicted values of $\{\gamma ,\GG\}$ 
for the kinetic gravity braiding model~\cite{Kimura:2010di}~($n=1,2,3$)
as purple boxes and the large $\GG$ model with $p=1$, derived in Appendix~\ref{sec:Large
Gamma model} as orange triangle.
Although the constraint from galaxy bispectrum on the gravitational growth index $\gamma$ is relatively
weaker than the expected constraints by galaxy power spectrum, the precise measurement
conducted by future galaxy surveys can constrain $\GG$ significantly. 
In particular, we can distinguish the models in which the expansion
history and the linear growth rate
are almost same as the fiducial mode but the different nonlinear evolution is given.

%%%%%%%%%%%%%%%%%%%%%%%%%%%%%%%%%%%%%%%%%%%%%%%%%%%%%%%%%%%%%%%%%%%%%%%%%
%=======================================================================%
\section{Conclusion}
\label{sec:conclusion}
%=======================================================================%
%%%%%%%%%%%%%%%%%%%%%%%%%%%%%%%%%%%%%%%%%%%%%%%%%%%%%%%%%%%%%%%%%%%%%%%%%

In this paper, we have discussed the potential power of the bispectrum of biased objects
as a possible new probe to test the theory of gravity beyond the linear-order perturbation.
To investigate the impact of the galaxy bispectrum,
%we have firstly performed the examination of
we have performed the generalization of
the redshift-space galaxy bispectrum
%for the density contrast generalized
to the wider class
of gravity theory based on the standard cosmological perturbation theory.
Since the modification of gravity theory typically changes the clustering property of large-scale structure,
measuring the galaxy bispectrum induced by the late-time nonlinear gravitational evolution of the density fluctuations
can be used to test the gravity theory through the evolution of the
linear growth rate and the second-order kernels.
Among them, in order to
focus on the time-evolving coefficient $\lambda$ in the second-order
kernel, we have introduced the second-order index $\GG$ 
defined in Eq.~\eqref{eq:lambda-Gamma def}, as a good candidate to catch
the higher-order nature of modified gravity theory.
We analytically obtained the expression of $\GG$ in terms of the parameters characterizing the growth history of the Universe 
[Eq.~\eqref{eq:Gamma estimation}]
as well as the gravitational growth index~$\gamma$.
We then found that the second-order index $\GG$
%growth index of $\lambda$
can carry new information about the growth of structure that is not
included in the linear perturbation theory.

As a specific model of modified gravity theory, we have applied the result to the most general scalar-tensor theory
with second-order equations of motion, namely the Horndeski theory. 
There are models in which the Hubble expansion and the gravitational growth index are completely same as the $\Lambda$CDM model
but the different value of $\GG$ can be obtained.
Finally, we have performed the Fisher matrix analysis to show that the future precise measurements of
galaxy bispectrum can significantly constrain $\GG$ as well as the
equation-of-state parameter and the gravitational growth index.

In this paper, we have made several simplified assumptions.
We have considered only the constant $\gamma$ and $\GG$ as the asymptotic values in high redshifts.
Although many modified gravity theories are known to be well described by constant $\gamma$ and $\GG$,
in some cases, it would be useful to consider the time-dependence of them.
Particularly we found that the time-dependence of $\GG$ generally
becomes large in lower redshifts.
When we consider a near-future survey that covers a comparative low-redshift depth, the effect of 
the time-dependence should be included to model the growth-history of the Universe precisely.
We hope to address these issues in more realistic situations.

%%%%%%%%%%%%%%%%%%%%%%%%%%%%%%%%%%%%%%%%%%%%%%%%%%%%%%%%%%%%%%%%%%%%%%%%%
%=======================================================================%
\acknowledgments
%=======================================================================%
%%%%%%%%%%%%%%%%%%%%%%%%%%%%%%%%%%%%%%%%%%%%%%%%%%%%%%%%%%%%%%%%%%%%%%%%%
%************************* Acknowledgments ******************************%

This work was supported in part by Grant-in-Aid from
the Ministry of Education, Culture, Sports, Science and Technology (MEXT) of Japan,  Nos.~17K14304 (D.~Y), 
15K17659, 15H05888, and 16H01103 (S.~Y.),
15K17646, 17H01110 (H.~T.).
We thank Kazuhiro Yamamoto, Tsutomu Kobayashi, and Shin'ichi Hirano for useful comments.

\appendix

%%%%%%%%%%%%%%%%%%%%%%%%%%%%%%%%%%%%%%%%%%%%%%%%%%%%%%%%%%%%%%%%%%%%%%%%%
%=======================================================================%
\section{Coefficients} \label{sec:coefficients}
%=======================================================================%
%%%%%%%%%%%%%%%%%%%%%%%%%%%%%%%%%%%%%%%%%%%%%%%%%%%%%%%%%%%%%%%%%%%%%%%%%

In this Appendix, we summarize the definitions of the coefficients in the field equation presented in Sec.~\ref{sec:Horndeski theory of gravity}.
The coefficients in the linear part are defined by
\al{
	&A_0=\frac{\dot\Theta}{H^2}+\frac{\Theta}{H}+{\cal F}_{\rm T}-2{\cal G}_{\rm T}-2\frac{\dot{\cal G}_{\rm T}}{H}+\frac{\rho_{\rm m}}{2H^2}
	\,,\\
	&A_1=\frac{\dot{\cal G}_{\rm T}}{H}+{\cal G}_{\rm T}-{\cal F}_{\rm T}
	\,,\\
	&A_2={\cal G}_{\rm T}-\frac{\Theta}{H}
	\,.
}
where
\al{
	&{\cal F}_{\rm T}=2\Bigl[G_4-X\left(\ddot\phi G_{5X}+G_{5\phi}\right)\Bigr]
	\,,\\
	&{\cal G}_{\rm T}=2\Bigl[G_4-2XG_{4X}-X\left(H\dot\phi G_{5X}-G_{5\phi}\right)\Bigr]
	\,,\\
	&\Theta =-\dot\phi XG_{3X}+2HG_4-8HXG_{4X}-8HX^2G_{4XX}+\dot\phi G_{4\phi}+2X\dot\phi G_{4\phi X}
	\notag\\
	&\quad\quad\quad\quad
		-H^2\dot\phi\left( 5XG_{5X}+2X^2G_{5XX}\right)
		+2HX\left(3G_{5\phi}+2X G_{5\phi X}\right)
	\,.
}
The coefficients in the higher-order parts are given by
\al{
	&B_0=\frac{X}{H}\biggl[
			\dot\phi G_{3X}+3\left(\dot X+2HX\right) G_{4XX}+2X\dot XG_{4XXX}-3\dot\phi _{4\phi X}+2\dot\phi XG_{4\phi XX}
	\notag\\
	&\quad\quad
			+\left(\dot H+H^2\right)\dot\phi G_{5X}
			+\dot\phi \left(2H\dot X+(\dot H+H^2)X\right) G_{5XX}+H\dot\phi X\dot XG_{5XXX}
	\notag\\
	&\quad\quad
			-2\left(\dot X+2HX\right) G_{5\phi X}
			-\dot\phi XG_{5\phi\phi X}-X\left(\dot X-2HX\right) G_{5\phi XX}
		\biggr]
	\,,\\
	&B_1=2X\biggl[G_{4X}+\ddot\phi (G_{5X}+XG_{5XX})-G_{5\phi}+XG_{5\phi X}\biggr]
	\,,\\
	&B_2=-2X\left( G_{4X}+2XG_{4XX}+H\dot\phi G_{5X}+H\dot\phi XG_{5XX}-G_{5\phi}-XG_{5\phi X}\right)
	\,,\\
	&B_3=H\dot\phi XG_{5X}
%	\,,\\
%	&C_0=2X^2G_{4XX}+\frac{2}{3}X^2\left(2\ddot\phi G_{5XX}+\ddot\phi XG_{5XXX}-2G_{5\phi X}+XG_{5\phi XX}\right)
%	\,,\\
%	&C_1=H\dot\phi X\left(G_{5X}+XG_{5XX}\right)
	\,.
}

%%%%%%%%%%%%%%%%%%%%%%%%%%%%%%%%%%%%%%%%%%%%%%%%%%%%%%%%%%%%%%%%%%%%%%%%%

%%%%%%%%%%%%%%%%%%%%%%%%%%%%%%%%%%%%%%%%%%%%%%%%%%%%%%%%%%%%%%%%%%%%%%%%%
%=======================================================================%
\section{Large $\GG$ model}
\label{sec:Large Gamma model}
%=======================================================================%
%%%%%%%%%%%%%%%%%%%%%%%%%%%%%%%%%%%%%%%%%%%%%%%%%%%%%%%%%%%%%%%%%%%%%%%%%

In this section, we try to construct the model in which there is observably large deviation 
for the second order index
%of the second-order kernel
$\GG$ but the gravitational growth index $\gamma$
remains the close value to the standard one.
To proceed the model building, we consider the subclass of Horndeski theory with the shift-symmetry, 
namely $K_\phi =G_{i,\phi}=0$, and $G_5=0$.
In this setup, the EFT parameters reduce to
\al{
	&M^2=2\left( G_4-2XG_{4X}\right)
	\,,\\
	&HM^2\alpha_{\rm M}=-2\dot X\left( G_{4X}+2XG_{4XX}\right)
	\,,\label{eq:alpha_M}\\
	&M^2\alpha_{\rm T}=4XG_{4X}
	\,,\label{eq:alpha_T}\\
	&HM^2\alpha_{\rm B}=2\dot\phi XG_{3X}+8HX\left( G_{4X}+2XG_{4XX}\right)
	\,,\\
	&H^2M^2\alpha_{\rm K}
		=2X\left( K_X+2XK_{XX}\right)
			+12\dot\phi HX\left(G_{3X}+XG_{3XX}\right)
	\notag\\
	&\quad\quad\quad\quad\quad
			+12H^2X\left(G_{4X}+8XG_{4XX}+4X^2G_{4XXX}\right)
	\,,\label{eq:alpha_K}\\
	&M^2\alpha_{\rm V1}=-2X\left( G_{4X}+2XG_{4XX}\right)
	\,,\\
	&M^2\alpha_{\rm V2}=0
	\,.\label{eq:alpha_V2}
}
The gravitational field equations are given by \cite{Kimura:2011dc}
\al{
	&{\cal E}=-\rho_{\rm m}
	\,,\ \ \ 
	{\cal P}=0
	\,,
}
where
\al{
	&{\cal E}=2XK_X-K+6H\dot\phi XG_{3X}-6H^2\Bigl[G_4-4X(G_{4X}+XG_{4XX})\Bigr]
	\,,\\
	&{\cal P}=K-\dot\phi\dot XG_{3X}+2(3H^2+2\dot H)G_4-4\Bigl[(3H^2+2\dot H)X+H\dot X\Bigr]G_{4X}-8HX\dot XG_{4XX}
	\,.
}
Here $\rho_{\rm m}$ is the nonrelativistic matter energy density.
Assuming the presence of the shift symmetry of the scalar field, $\phi\to\phi +{\rm const}.$,
we obtain the conservation equation of the Noether current, which is given by
\al{
	\dot J+3HJ=0
	\,,
}
where the Noether current is defined as
\al{
	&J=\dot\phi K_X+6HXG_{3X}+6H^2\dot\phi\left( G_{4X}+2XG_{4XX}\right)
	\,.\label{eq:J def}
}
We clearly find that its solution is given by $J\propto 1/a^3$, implying that
$J$ approaches zero as the universe expands.
Therefore, we take $J=0$ as the simplest attractor solution throughout the paper.
With the attractor condition and the $\alpha$ functions defined in Eqs.~\eqref{eq:alpha_M}-\eqref{eq:alpha_V2},
the gravitational equations of motion can become the more suggestive form:
\al{
	&3M^2H^2=\rho_{\rm m}-K% by SY -J\dot\phi
	\,,\label{eq:Friedmann1}\\
	&M^2\left( 2\dot H+3H^2\right)=-K+\frac{1}{2}\frac{\dd\ln X}{\dd\ln a}H^2M^2\alpha_{\rm B}
	\,,\label{eq:Friedmann2}
}
which immediately implies that the corresponding dark energy model is given by
\al{
	&\rho_{\rm DE}=-K% by SY -J\dot\phi
	\,,\\
	&p_{\rm DE}=K-\frac{1}{2}\frac{\dd\ln X}{\dd\ln a}H^2M^2\alpha_{\rm B}
	\,.
}
The attractor solution $J=0$ provides the conditions for the $\alpha$ functions:
\al{
	&J\dot\phi =2XK_X+3H^2M^2\left(\alpha_{\rm B}+2\alpha_{\rm V1}\right)=0
	\,,\label{eq:J eq1}\\
	&\dot J\dot\phi =M^2H^3\Biggl(\frac{1}{2}\frac{\dd\ln X}{\dd\ln a}\alpha_{\rm K}+3\frac{\dd\ln H}{\dd\ln a}\alpha_{\rm B}\Biggr) =0
	\,.\label{eq:J eq2}
}
Hence, the equation-of-state parameter for the dark energy can be written as
\al{
	w_{\rm DE}=-1+\frac{1}{1-\widetilde\Omega_{\rm m}}\frac{\dd\ln H}{\dd\ln a}\frac{\alpha_{\rm B}^2}{\alpha_{\rm K}}
	\,.\label{eq:EoS reduced}
}
Substituting \eqref{eq:EoS reduced} into Eqs.~\eqref{eq:dLogHdLoga eq} and \eqref{eq:dLogOmegamdLoga eq}, we then have
\al{
	&\frac{\dd\ln H}{\dd\ln a}
		=-\frac{3}{2}\widetilde\Omega_{\rm m}\frac{1}{1+\alpha_{\rm BK}}
	\,,\label{eq:dLogHdLoga}\\
	&\frac{\dd\ln\widetilde\Omega_{\rm m}}{\dd\ln a}
		=-3\left( 1-\widetilde\Omega_{\rm m}\right) \SY{-} 3\widetilde\Omega_{\rm m}\frac{\alpha_{\rm BK}}{1+\alpha_{\rm BK}}-\alpha_{\rm M}
	\,,\label{eq:dOmegamdLoga}
}
where $\alpha_{\rm BK} =3\alpha_{\rm B}^2/2\alpha_{\rm K}$ and
Eq.~\eqref{eq:EoS reduced} can be given by
\al{
	w_{\rm DE}=-1 \SY{-}\frac{\widetilde\Omega_{\rm m}}{1-\widetilde\Omega_{\rm m}}\frac{\alpha_{\rm BK}}{1+\alpha_{\rm BK}}
	\,.\label{eq:w_DE}
}
It leads that the asymptotic value of $w_{\rm DE}$ at the early stage of the Universe is obtained as
\al{
	w^{(0)}=-1-\frac{3c_{\rm B}^2}{2c_{\rm K}}
	\,.
}

Using these solutions, we would like to construct the explicit model to realize the large $\GG$ with small deviation of $\gamma$.
Let us assume the following form of the Horndeski functions (see also \cite{DeFelice:2011bh,DeFelice:2011aa} for the extended Galileon model) :
\al{
	&K=-c_2M_2^4\left(\frac{X}{M_2^4}\right)^{p_2}
	\,,\ \ 
	G_3=c_3M_3\left(\frac{X}{M_3^4}\right)^{p_3}
	\,, \ \ 
	G_4=\frac{M_*^2}{2}-c_4M_4^2\left(\frac{X}{M_4^4}\right)^{p_4}
	\,.
}
We search for a tracker solution, which is characterized by the condition 
\al{
	H\dot\phi^{2q}={\rm const}.%=H_{\rm dS}\dot\phi_{\rm dS}^{2q}
	\,,
}
where $q$ is assumed to be real constant.
%, $H_{\rm dS}$ and $\dot\phi_{\rm dS}$ denote the constant parameters characterizing a de Sitter solution.
When we choose the following powers, all terms in Eq.~\eqref{eq:J def} are proportional to $\dot\phi^{2p}$: 
\al{
	p_2=p\,,\ \ \ 
	p_3=p+q-\frac{1}{2}\,,\ \ \ 
	p_4=p+2q
	\,.
}
With them, the attractor condition $J=0$ gives the relation between $c_i$ and $M_i$.
%\al{
%	c_2=\frac{1}{p}3\times 2^{-1-2q}H_{\rm dS}\dot\phi_{\rm dS}^{2q}M_2^{-4(1-p)}
%			\biggl[
%				2^{\frac{1}{2}+q}c_3M_3^{3-4p-4q}(-1+2p+2q)-4c_4H_{\rm dS}\dot\phi_{\rm dS}^{2q}M_4^{2(1-2p-4q)}(-1+2p+4q)
%			\biggr]
%	\,.
%}
Eqs.~\eqref{eq:J eq1} and \eqref{eq:J eq2} yield the relation between $\alpha$ functions:
\al{
	\alpha_{\rm B}=2p\left( 1-\widetilde\Omega_{\rm m}\right) -2\alpha_{\rm V1}
	\,,\ \ 
	\alpha_{\rm K}=6q\alpha_{\rm B}
	\,.
}
Moreover, Eqs.~\eqref{eq:alpha_M} and \eqref{eq:alpha_T} can be rewritten as
\al{
	&\alpha_{\rm M}=\frac{3}{2q}\widetilde\Omega_{\rm m}\frac{\alpha_{\rm V1}}{1-\alpha_{\rm BK}}
	\,,\ \ 
	\alpha_{\rm T}=\frac{2\alpha_{\rm V1}}{1-2p-4q}
	\,.
}
Combining these, we can write down the explicit form of $c_i$ in terms of $p$, $q$ and $c_{\rm B}$ as
\al{
	c_{\rm K}=6qc_{\rm B}
	\,,\ \ 
	c_{\rm M}=\frac{3}{2q}\left( p-\frac{1}{2}c_{\rm B}\right)
	\,,\ \ 
	c_{\rm T}=\frac{2}{1-2p-4q}\left( p-\frac{1}{2}c_{\rm B}\right)
	\,,\ \ 
	c_{\rm V1}=p-\frac{1}{2}c_{\rm B}
	\,,\ \ 
	c_{\rm V2}=0
	\,.
}
Here, we consider the negligible braiding case as a specific example,
that is $c_{\rm B}\to 0$.
In this case,
$c_{\rm K}$ gives no contributions to the density perturbations in the small scale limit and 
there is no dependence on $c_{\rm T}$ in the case of $c_{\rm B}\to 0$, as shown in the previous section.
Since $c_{\rm M}=3p/2q$ and $c_{\rm V1}=p$, the leading order index of the growth and the second-order kernel
can be rewritten in terms of $p$ and $q$. 
In particular, when we consider the large hierarchy $|p/q|\ll 1$, Eqs.~\eqref{eq:gamma0} 
and \eqref{eq:Gamma0} are given by
\al{
	\gamma\approx 0.545
	\,,\ \ \ 
	\GG \approx 0.005-0.151p
	\,.
}
We found that there is one parameter family of the model in which the
gravitational growth index $\gamma$ is the same as the standard $\Lambda$CDM+GR value 
while the second-order index~$\GG$ can have the large deviation from the standard one.

\end{document}